\documentclass[12pt,epsfig]{article}
\newcommand{\ba}{\begin{array}{c}}
\newcommand{\baz}{\begin{array}{cc}}
\newcommand{\bad}{\begin{array}{ccc}}
\newcommand{\bav}{\begin{array}{cccc}}
\newcommand{\ea}{\end{array}}

\usepackage{amssymb} 
\usepackage{epsfig}
\usepackage{float}
\newcommand{\be}{\begin{equation}}
\newcommand{\ee}{\end{equation}}
\newcommand{\bea}{\begin{eqnarray}}
\newcommand{\eea}{\end{eqnarray}}
\begin{document}

\begin{center}
\bf {A recursive parameterisation of unitary matrices}
\end{center}

\begin{center}
C. Jarlskog 
\end{center}

\begin{center}
{\em Division of Mathematical Physics\\
LTH, Lund University\\
Box 118, S-22100 Lund, Sweden}
\end{center}

\begin{abstract}

A simple recursive scheme for parameterisation of 
n-by-n unitary matrices is presented.
The n-by-n matrix is expressed as a product containing the
(n-1)-by-(n-1) matrix and 
a unitary matrix that contains the additional parameters needed to go from
n-1 to n. The procedure is repeated to obtain recursion 
formulas for n-by-n unitary matrices.

\end{abstract}

\section{The parameterisation}

It has been known since a long time that unitary transformations
play a central role in physics. An excellent example is Wigner's paper of 1939 
\cite{wigner39} which has had a great
impact in the development of physics and is still important.  

It is also a known fact that
a general n-by-n unitary matrix $X^{(n)}$ may be expressed 
as a product of three unitary matrices,
\begin{equation}
X^{(n)} = \Phi^{(n)}(\vec{\alpha}) V^{(n)} \Phi^{(n)}(\vec{\beta})
\label{defx}
\end{equation}
where the matrices $\Phi$ are diagonal unitary matrices, 
\begin{equation} 
\Phi^{(n)}(\vec{\alpha}) = 
\left( \begin{array}{ccccc}
e^{i\alpha_1}&&&& \\
& e^{i\alpha_2}&&& \\
&& .&&\\
&&& . & \\
&&&& e^{i\alpha_n}
\end{array} \right)
\end{equation}
$\Phi(\vec{\beta})$ is defined analogously; the $\alpha$'s and
$\beta$'s being real. The matrix $X^{(n)}$ has $n^2$ real
parameters. In the following, for simplicity, the word parameter stands
for real parameter. The quantities $\vec{\alpha}$
and $\vec{\beta}$ take care of $2n-1$ parameters of $X^{(n)}$
because only the sums $\alpha_i+ \beta_j$ 
enter, where $i$ and $j$ run from $1$ to $n$. The remaining
$(n-1)^2$ parameters reside in the non-trivial matrix $V^{(n)}$ which is the 
subject of this study.

We start by putting $V^{(1)} =1$, whereby $X^{(1)}= 
e^{i(\alpha_1 + \beta_1)}$ is the
most general one-by-one unitary "matrix".   
For $n \ge 2$, we write the matrix $V^{(n)}$  
in the form
\begin{equation}
V^{(n)} = \left(  \begin{array}{cc}  V^{(n-1)}  + 
(1-c_n)\vert A^{(n-1)}> < B^{(n-1)} \vert ~~ & s_n
\vert A^{(n-1)} > \\ s_n <B^{(n-1)} \vert & c_n   \end{array} \right)   
\label{defvn} 
\end{equation}
Here we have introduced an angle denoted by $\theta_n$ and 
have used the common notation
$s_n = sin\theta_n$, $c_n= cos\theta_n$. 
The complex vectors "$A^{(n-1)}$" and "$B^{(n-1)}$"
have each n-1 components, i.e.,
\begin{equation}
\vert A^{(n-1)}> = \left( \begin{array}{c}
a_1^{(n-1)} \\ a_2^{(n-1)} \\ .\\.\\ a_{n-1}^{(n-1)} \end{array}
\right),~~~~
\vert B^{(n-1)}> = \left( \begin{array}{c}
b_1^{(n-1)} \\ b_2^{(n-1)} \\ .\\.\\ b_{n-1}^{(n-1)} \end{array}
\right)
\label{ab}
\end{equation}
Furthermore,
\begin{equation}
< B^{(n-1)} \vert = 
(b_1^{(n-1)\star}, b_2^{(n-1)\star},~.~., b_{n-1}^{(n-1)\star})
\end{equation}
and
\begin{equation} 
(\vert A^{(n-1)} > < B^{(n-1)} \vert)_{ij} 
\equiv a^{(n-1)}_i b^{(n-1)\star}_j
\end{equation}
$A^{(n-1)}$ and $B^{(n-1)}$ are not arbitrary but are required to satisfy 
the conditions
\begin{equation}
<A^{(n-1)} \vert A^{(n-1)}> =1, ~~~~ 
\vert B^{(n-1)}> = -V^{(n-1)\dagger} \vert A^{(n-1)}>
\label{relab}  
\end{equation}
whereby
\begin{equation}
<B^{(n-1)} \vert B^{(n-1)}> =1, ~~~~ 
\vert A^{(n-1)}> = -V^{(n-1)} \vert B^{(n-1)}>  
\end{equation}

We can easily check that if the matrix $V^{(n-1)}$,
in Eq.(\ref{defvn}), is unitary
so is $V^{(n)}$. In order for
$V^{(n)}$ to be the most general n-by-n unitary
matrix, modulus the phase matrices $\Phi$, it must have the
required number of parameters.  
The vector $A^{(n-1)}$, having $n-1$ complex components, would seem to represent
$2(n-1)$ parameters. But it has only $2(n-2)$ because it is
normalised and its overall phase can be absorbed into the
matrices $\Phi$, i.e., the transformation
\begin{equation}
\vert A^{(n-1)}>~ \rightarrow e^{i \eta} \vert A^{(n-1)}>
\end{equation}
yields
\begin{equation}
\vert B^{(n-1)}> ~\rightarrow e^{i \eta} \vert B^{(n-1)} >
\end{equation} 
and 
\begin{equation}
V^{(n)} \rightarrow \Phi (0,0,...,e^{-i \eta}) V^{(n)}
\Phi (0,0,...,e^{i \eta})
\end{equation}
The parameter counting, therefore, goes as follows. On the LHS of 
Eq.(\ref{defvn})
we need to have $(n-1)^2$ parameters. On the RHS, we have $(n-2)^2$ from
$V^{(n-1)}$ and $2(n-2)$ from the vector $A^{(n-1)}$
Thus, together with the angle $\theta_n$, the number of
parameters is
$(n-2)^2 + 2 (n-2) +1$ which equals $(n-1)^2$ as required.

We may use the relation (\ref{relab}) between $A^{(n-1)}$ and $B^{(n-1)}$
to rewrite the matrix $V^{(n)}$ in  
terms of either $A^{(n-1)}$ or $B^{(n-1)}$. In terms of $A^{(n-1)}$, we have 
\begin{eqnarray}
V^{(n)} &=&  
\left( \begin{array}{cc} 1-(1-c_n))\vert A^{(n-1)}><A^{(n-1)} 
\vert & ~s_n \vert A^{(n-1)}> \\
-s_n < A^{(n-1)} \vert & ~c_n \end{array} \right)
\left( \begin{array}{cc}
V^{(n-1)} & 0\\
0& 1
\end{array} \right) \nonumber \\
& \equiv & A_{n,n-1}
\left( \begin{array}{cc}
V^{(n-1)} & 0\\
0& 1
\end{array} \right) 
\label{recursiona}
\end{eqnarray}
While writing the matrix in terms of $B^{(n-1)}$ yields 
\begin{eqnarray}
V^{(n)} & = &\left( \begin{array}{cc}
V^{(n-1)} & 0\\
0& 1
\end{array} \right) 
\left( \begin{array}{cc} 1-(1-c_n))\vert B^{(n-1)}><B^{(n-1)} 
\vert & ~-s_n \vert B^{(n-1)}> \\
s_n < B^{(n-1)} \vert & ~c_n \end{array} \right) \nonumber \\
& \equiv & \left( \begin{array}{cc} 
V^{(n-1)} & 0\\
0& 1
\end{array} \right) B_{n,n-1}
\label{recursionb}
\end{eqnarray}
These relations allow for a systematic construction of
unitary matrices order by order.
By repeating the above procedure for the matrix $V^{(n-1)}$
in terms of $A^{(n-2)}$ and $B^{(n-2)}$, and following down
the chain we find the recursion formulas that we are looking for,
\begin{equation}
V^{(n)}=A_{n,n-1}A_{n,n-2}...A_{n,2}A_{n,1}
\label{vna}
\end{equation}
\begin{equation}
V^{(n)}= B_{n,1}B_{n,2}... B_{n,n-2}B_{n,n-1}
\label{vnb}
\end{equation}
The matrices $A_{n,n-1}$ and $B_{n,n-1}$ were 
previously defined in Eqs.(\ref{recursiona}) and
(\ref{recursionb}). For $j < n-1$ we have 
\begin{equation}
A_{n,j} =
\left( \begin{array}{cc}
 \left( \begin{array}{cc} 1-(1-c_{j+1}))\vert A^{(j)}><A^{(j)} 
\vert & ~s_{j+1} \vert A^{(j)}> \\
-s_{j+1} < A^{(j)} \vert & ~c_{j+1} \end{array} \right)&~~0 \\
0~~ & I_{n-j-1} \end{array} \right)
\label{anj}
\end{equation}
\begin{equation}
B_{n,j} =
\left( \begin{array}{cc}
 \left( \begin{array}{cc} 1-(1-c_{j+1}))\vert B^{(j)}><B^{(j)} 
\vert & ~-s_{j+1} \vert B^{(j)}> \\
s_{j+1} < B^{(j)} \vert & ~c_{j+1} \end{array} \right)&~~0 \\
0~~ & I_{n-j-1} \end{array} \right)
\label{bnj}
\end{equation}
Here $I_{n-j-1}$ is the unit matrix of order $n-j-1$.
The two unitary matrices $A_{n,j}$ and $B_{n,j}$ are
related by 
\begin{equation}
A_{n,j} = \left( \begin{array}{cc} V^{(j)} & 0 \\
0 & I_{n-j} \end{array}\right)     B_{n,j}
\left( \begin{array}{cc} V^{(j)\dagger} & 0 \\
0 & I_{n-j} \end{array}\right)
\end{equation}

\section{Simple examples}
The simplest case is $n=2$ for which we take $\vert A^{(1)}> = 1$
whereby $\vert B^{(1)}> =-1$. Using $V^{(1)} =1$
we obtain, from Eqs.(\ref{recursiona}) and (\ref{recursionb}) 
\begin{equation}
V^{(2)} = A_{2,1} = B_{2,1} = \left( \begin{array}{cc} c_2& s_2\\
-s_2 & c_2 \end{array} \right) 
\label{rot2}   
\end{equation}
This is the familiar rotation matrix $R_2(\theta_2)$, 
$\theta_2$ being the rotation angle in two dimensions. 

The next simplest case is $n=3$ for which we may either use the "mixed form"
or the pure forms. For the mixed form we have
\begin{equation}
V^{(3)}= \left( \begin{array}{cc} R_2(\theta) + 
(1-c_3)\vert A^{(2)}> < B^{(2)} \vert ~~ & s_3
\vert A^{(2)} > \\ s_3 <B^{(2)} \vert & c_3   \end{array} \right)  
\end{equation}
Here $R_2(\theta)$ is again the rotation matrix in Eq.(\ref{rot2}) and
we may put
\begin{equation}
\vert A^{(2)}> = \left( \begin{array}{c} a_1 \\ a_2 \end{array} \right), ~~
\vert B^{(2)}> = \left( \begin{array}{c} b_1 \\ b_2 \end{array} \right)
\end{equation}
From Eq.(\ref{relab}) follows that  
\begin{equation}
\vert A^{(2)}> = -R_2(\theta) \vert B^{(2)}>, ~~~~
\vert B^{(2)}> = -R_2(-\theta) \vert A^{(2)}> 
\end{equation}
Hence these vectors represent two parameters, for example 
\begin{equation}
\vert A^{(2)}> = \left( \begin{array}{c} cos\gamma \\ 
sin\gamma e^{i\delta}  \end{array} \right)
\label{defa2}
\end{equation}
where $\gamma$ and $\delta$ are real.
The pure forms are also obtained very simply, for example

\begin{equation}
V^{(3)}= \left( \begin{array}{cc} 1-
(1-c_3)\vert A^{(2)}> < A^{(2)} \vert ~~ & s_3
\vert A^{(2)} > \\ -s_3 <A^{(2)} \vert & c_3   \end{array} \right)
\left( \begin{array}{cc}
 R_2(\theta_2)   & 0\\
0& 1
\end{array} \right)  
\end{equation}
where $A^{(2)}$ is as defined in Eq.(\ref{defa2}).
Evidently, depending on the application one has in mind some choices may be
more convenient than others. This is demonstrated in 
Refs.\cite{cejacab} and \cite{lepcab} which deal with the 
so called quark and lepton mixing matrices. The essential point is that
$V^{(3)}$ is described in a rather simple fashion
by four parameters as it should be. 
For the case of $n=4$ we could, for example, take $A^{(3)}$ to be
\begin{equation}
\vert A^{(3)}> = \left( \begin{array}{c} cos\rho \\ 
sin\rho cos\sigma e^{i\delta_1} \\ 
sin\rho sin\sigma e^{i\delta_2} \end{array} \right)
\end{equation}
where $\rho$, $\sigma$, $\delta_1$ and $\delta_2$ are the four 
parameters needed to define the most general $A^{(3)}$.  

In principle, the above recursive procedure may "easily" 
be extended to much larger $n$ with the help of computers.

\end{document}